\documentclass[preprint,12pt,sort&compress]{elsarticle}

\usepackage[caption=false]{subfig}
\usepackage{amssymb,amsfonts,graphicx,amsmath,setspace,bm}
\usepackage[retainorgcmds]{IEEEtrantools}
\journal{arxiv.org}

\begin{document}

\begin{frontmatter}
\title{Translocation through a narrow pore under a pulling force}
\author{Mohammadreza Niknam Hamidabad}
\author[mysecondaryaddress]{Rouhollah Haji Abdolvahab\corref{mycorrespondingauthor}}
\cortext[mycorrespondingauthor]{Corresponding author}
\ead{rabdolvahab@gmail.com}
\address{Physics Department, Iran University of Science and Technology (IUST), 16846-13114, Tehran, Iran.}

\begin{abstract}
We employ a three-dimensional molecular dynamics to simulate translocation of a polymer through a nanopore driven by an external force. The translocation is investigated for different four pore diameters and two different external forces. In order to see the polymer and pore interaction effects on translocation time, we studied 9 different interaction energies. Moreover, to better understand the simulation results we investigate polymer center of mass, shape factor and the monomer spatial distribution through the translocation process.

Our results unveil that while increasing the polymer-pore interaction energy slows down the translocation, expanding the pore diameter, makes the translocation faster. The shape analysis of the results reveals that the polymer shape is very sensitive to the interaction energy. In great interactions, the monomers come close to the pore from both sides. As a result, the translocation becomes fast at first and slows down at last.

\end{abstract}


\end{frontmatter}

\section{Introduction}
\label{intro}
Biopolymers translocation through nanopores is a critical and ubiquitous process in both biology and biotechnology. This leads to extensive and comprehensive studies over the past few decades. Undoubtedly, the study of the translocation of a polymer through nanopores can be considered as one of the most active fields of research in the whole soft matter physics \cite{muthuJCP99,kaifuJCP06,Al09,AbdPRE11,metzr14,15sean,16ren,17muthukumar}.
It should be noted that the importance of this process, polymer translocation (PT), is not limited to understanding its physical and biological dimensions, but also the essential technological applications, including DNA sequencing \cite{Nakane03,Abd08, Branton08, Liang15, Dekker18NN}, controlled drug delivery \cite{Tsutsui04,12chang}, gene therapy \cite{98Hanss,Tsutsui04, Wang07, Hepp16} and biological sensors \cite{Nakane03, 18Farimani}.

Moreover, the passage of biopolymers such as DNA and RNA through nuclear pore complexes \cite{01ribbeck, Meller03, 17Gallucci, 18Lai}, virus RNA injection into the host cell \cite{Inamdar06,18Chen} and passing proteins through the cell organelle membrane channels \cite{Al09} are some other biologic processes which have doubled the importance of this issue.

In the process, the biopolymer must overcome the entropy barrier \cite{SuPa96,muthuJCP99,muthuAnn07,11muthukumar}. Hence, the methods of PT  through the nanopores include the use of external force which is one of the most common methods used both in the laboratory and computational simulations \cite{Huopaniemi07,08lehtola,metzr14,15sean,16Magill,17Sun}. However, \textit{in vivo} PT driven by assisted proteins called chaperone is proposed \cite{A&M4,kaifujacs,AbdPRE11,Suhonen16,17Emamyari,AbdEPJ17}.

In the following simulation, we have used the polymer-mediated passage of polymer through nanopores driven by the external force. In this type of translocation, several parameters, such as the length and radius of the nanopore, the applied external force, and the friction coefficient of both Cis and Trans environments, are investigated \cite{kaifuJCP06MC,kaifuJCP06,Huopaniemi06,15Wang,18Ghosh}. In the meantime, one of the cases that are rarely investigated is the interaction energy (potential depth) between the nanopore wall and the polymer passing through it and its effect on the time of PT.

In this paper, we used a coarse-grained molecular dynamics method to simulate the translocation of the polymer through the nanopore in three dimensions. The simulation includes nine different interactions, three nanopore diameters, and two different external forces.

\section{Theory and simulation details}
\label{mm}

In the following 3D simulations, the polymer is modeled by a mass and spring chain in such a way that adjacent monomers have the nonlinear potential of FENE:

\begin{equation} \label{fene}
U_{FENE}= -\frac{1}{2} k R_{0}^{2} ln (1-\frac{ r^{2}}{ R_{0}^{2}}).
\end{equation} 

Here, r is the distance between two adjacent monomers, k, and $R_{0}$ are the spring constant and the maximum permissible spacing for adjacent monomers.

We employ the Lenard-Jones potential, equation \ref{lennard-jones}, to model the nanopore such that the cutoff radius of the nanopore interactions with the polymer is different from the other interactions.

\begin{equation} \label{lennard-jones}
U_{LJ}=\Bigg\{ \begin{array}{ll} 4 \epsilon \bigg[ \bigg(\frac{\sigma}{r}\bigg)^{12}  - \bigg(\frac{\sigma}{r}\bigg)^{6} \bigg] + \epsilon & r \leq r_{cut} \\ 
0 &  \text{otherwise} \end{array} 
\end{equation} 

in which $sigma$ is the diameter of each monomer, $\epsilon$ is the potential depth of the Lenard-Jones and $r_{cut}$ calls for the potential cutoff radius. 

We do the simulation employing the Langevin dynamics method. In this method, the following statement can be written for each monomer:

\begin{equation} \label{langevin}
m\ddot{r}=F_{i}^{C}+F_{i}^{F}+F_{i}^{R}
\end{equation}  

where m is the monomer mass. Moreover, the $F_{i}^{C}$, $F_{i}^{F}$, and $F_{i}^{R}$ are the conservative, frictional, and the random forces applied on the $i's$ monomer, respectively. The frictional forces are connected to the monomer's speed by the following equation:

\begin{equation} \label{monomervelocity}
F_{i}^{F}= - \xi V_{i}
\end{equation}  

in which $\xi$ is the frictional coefficient. One also can write for the conservative forces:

\begin{equation} \label{conservativeforce}
F_{i}^{C}=-\nabla (U_{LJ} + U_{FENE} ) + F_{external} 
\end{equation}

where the last term is the external force, excreted on the polymer through the nanopore and is defined as:

\begin{equation} \label{externalforce}
F_{external}=F \hat{x}
\end{equation}

in which, the direction of the force corresponds to the nanopore-axis and towards the Trans side.

\subsection{Simulation parameters}
\label{simpa}
The initial configuration of the system is such that the first monomer is at the end of a nanopore of length $6\sigma$ and with different diameters of $3, 4$ and $5\sigma$. We then place the remaining monomers close to their equilibrium position relative to each other, and in the front of the nanopore. It should be noted that the pore-axis is parallel to the x-axis. After placing the monomers, we give them the opportunity to achieve their equilibrium as a whole polymer. In this equilibration process, we fix a few monomers in the nanopore and allow the rest of the monomers, the polymer tail, to move freely until we reach the equilibrium. Afterward, the process of translocation begins. This part of the simulation lasts from about $20\%$, for the slowest, up to about $40\%$, for the fastest translocation, of each PT time through the nanopore. Here, we translocated the polymer for at least 1,500 times to reach a rather good time distribution.

In order to find the equilibrium point, we calculate the radius of gyration of the polymer through the time. The equilibration process continues until the changes in the radius of gyration becomes as small as $2\sigma$.

%
%

We calculate the time scale of the simulation using the $t_{LJ}$ which is \cite{kaifujacs}:

\begin{equation} \label{timescale}
t_{LJ}=(\frac{m \sigma^{2}}{\epsilon_{0}})^{\frac{1}{2}}
\end{equation}

We pick the forces from two different regions of strong and medium as the external force in the pore. The relation determining this regions for the average force is \cite{Huopaniemi07}:

\begin{equation} \label{forceequation}
\frac{k_{B}T}{\sigma N^{\nu}} \leq F \leq \frac{k_{B}T}{\sigma}
\end{equation}

in which $\nu$ is the Flory exponent, and $N$ stands for the total number of monomers. The magnitude of the strong and medium forces we employ in the simulation are $2\epsilon_{0}/\sigma$ and $1\epsilon_{0}/\sigma$, respectively ($\epsilon_{0}$ is defined below.).

\textbf{Simulation parameters:} The other simulation parameters include the cutoff radius for interactions of the nanopore and the polymer which is $2.5\sigma$ and in other interactions, between monomers and monomers and wall, are $2^{\frac{1}{6}}\sigma$. For the energy we use $\epsilon_{0}$ which is $\epsilon_{0}=1.2k_{B}T$, except the interaction between polymer and the nanopore which changed and are multiplies of $\epsilon_{0}$. Moreover, the friction coefficient is $\xi=0.7m/t_{LJ}$ and for the FENE potential, the spring constant is $k=30\epsilon_{0}/\sigma^{2}$ and the cutoff radius $R_{0}=1.5\sigma$ \cite{kaifujacs}.

\section{Results and analysis}
\label{result}

\begin{figure}[h]
  \includegraphics[width=\linewidth]{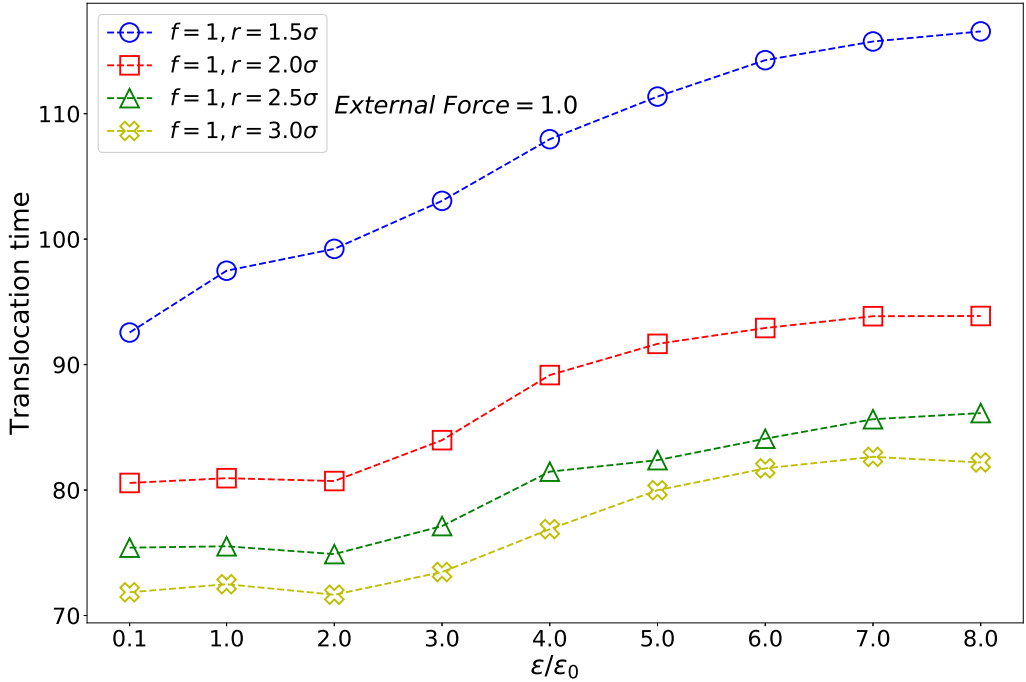}
  \caption{Translocation time versus energy for four different diameter in external force $f=1$}
  \label{timeepsilonf1}
\end{figure} 

\begin{figure}
  \includegraphics[width=\linewidth]{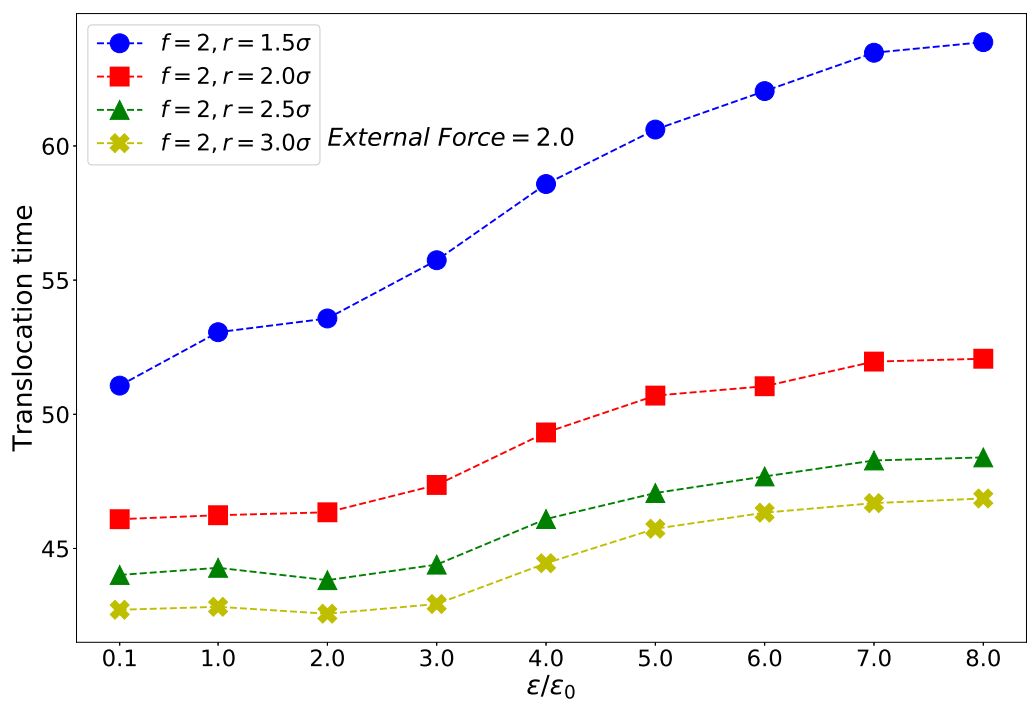}
  \caption{Translocation time versus energy for four different pore radii in external force $f=2$}
  \label{timeepsilonf2}
\end{figure}

\begin{figure}
  \includegraphics[width=\linewidth]{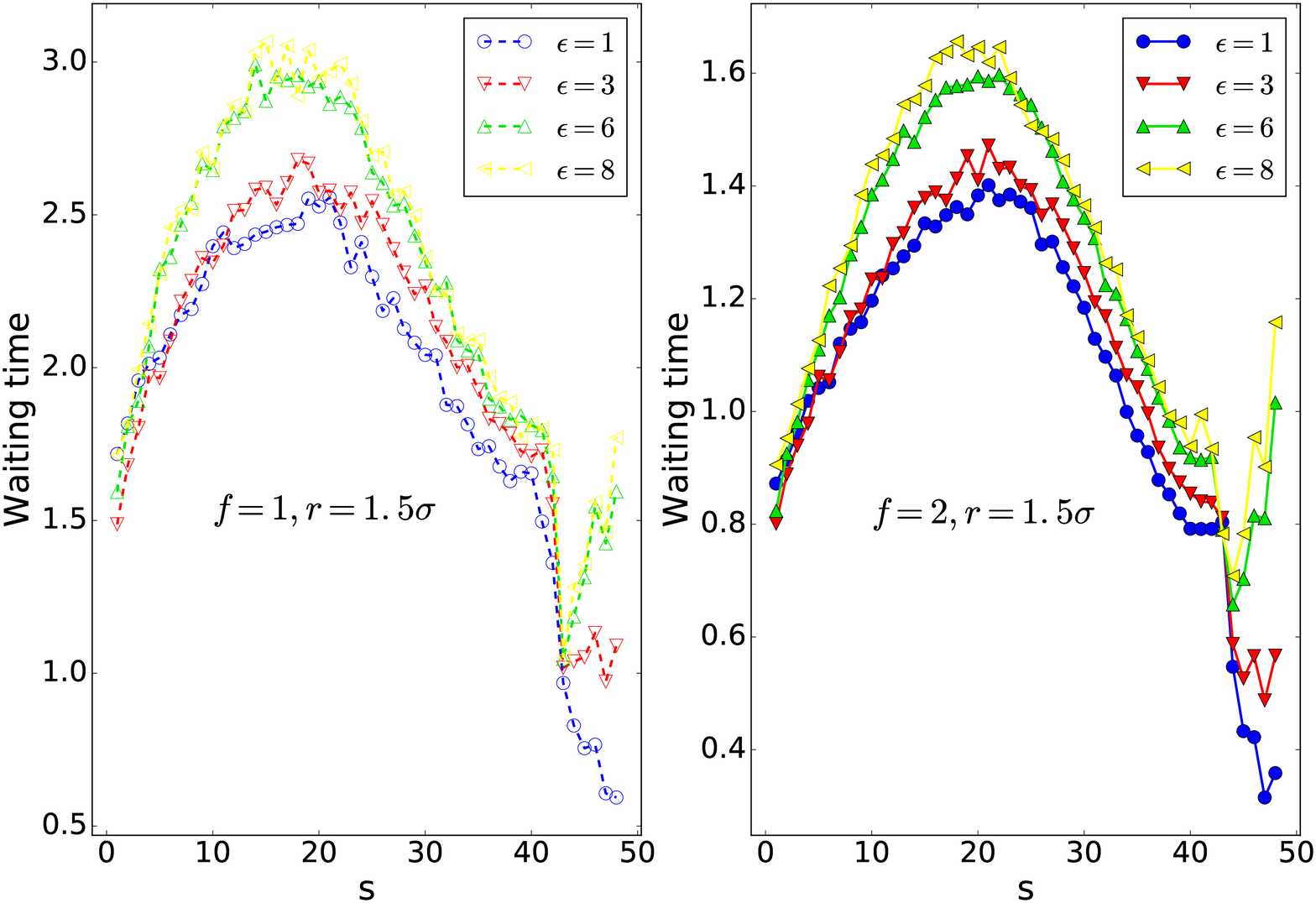}
  \caption{Mean waiting time of the polymer versus monomer number from a nanopore of radius $r=1.5\sigma$}
  \label{waiting1.5}
\end{figure} 
\begin{figure}
  \includegraphics[width=\linewidth]{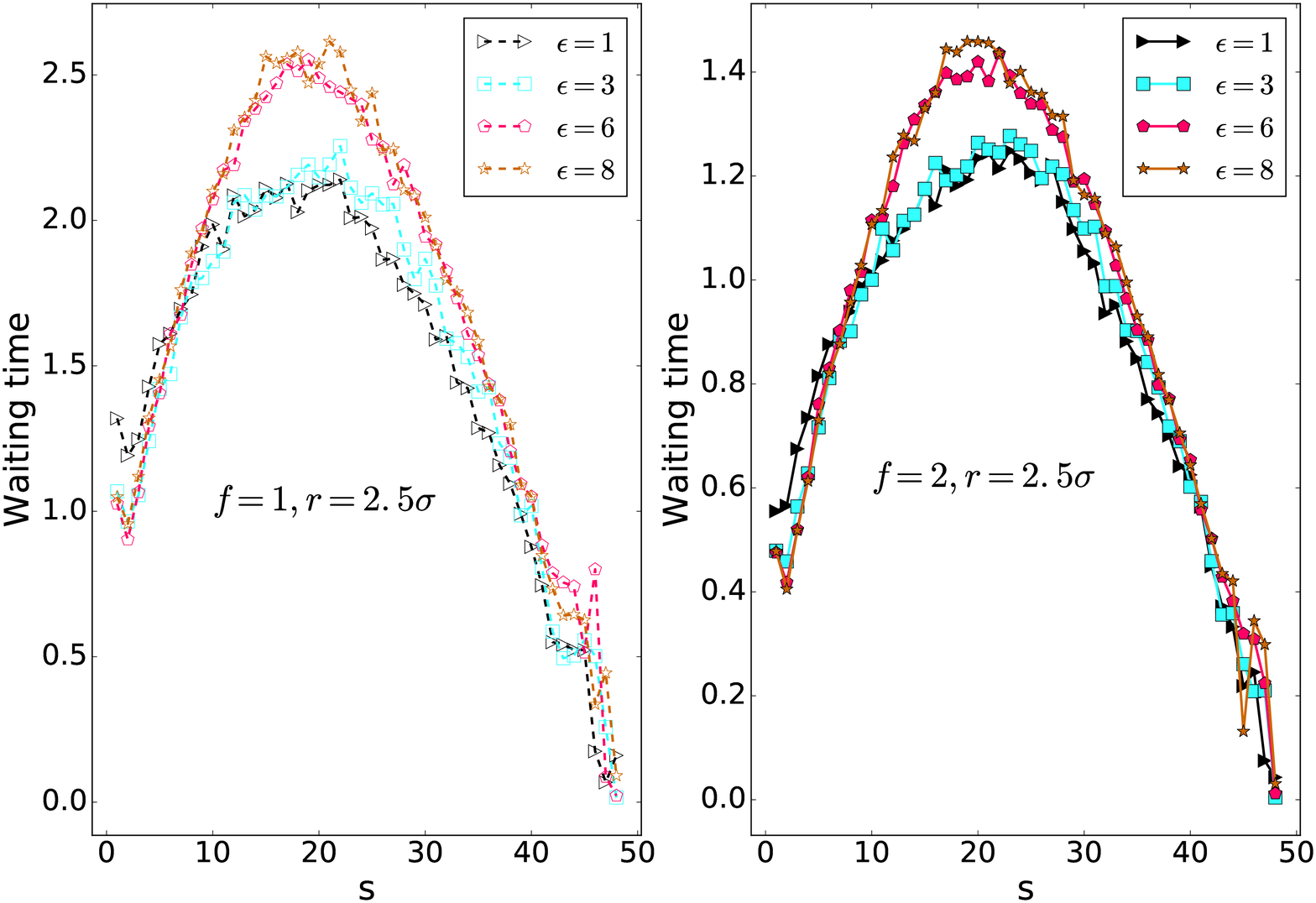}
  \caption{Mean waiting time of the polymer versus monomer number from a nanopore of radius $r=2.5\sigma$}
  \label{waiting2.5}
\end{figure} 
\begin{figure}
  \includegraphics[width=\linewidth]{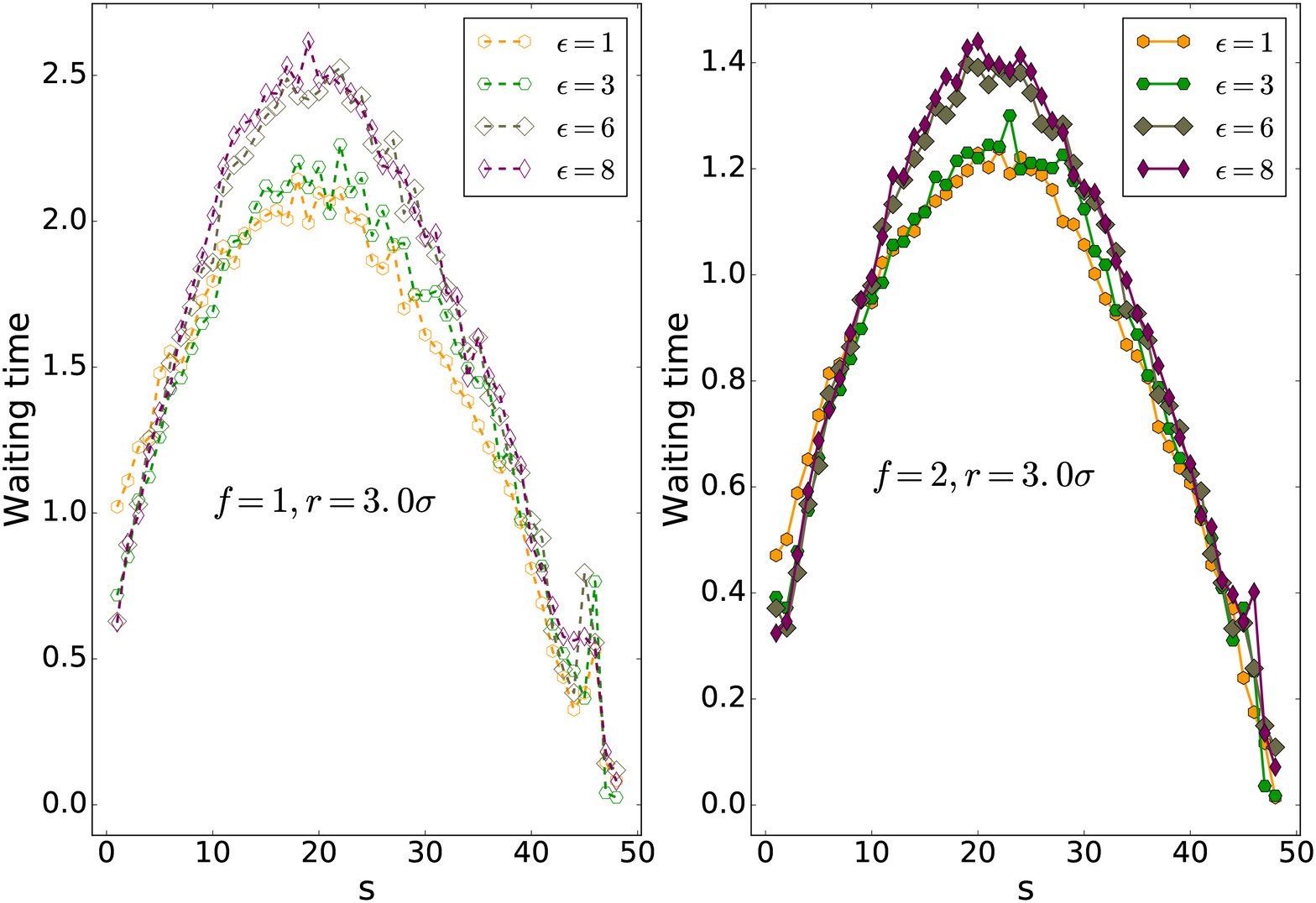}
  \caption{Mean waiting time of the polymer versus monomer number from a nanopore of radius $r=3.0\sigma$}
  \label{waiting3.0}
\end{figure} 

Translocation time of the polymer versus polymer and pore interaction energy is plotted in the figure \ref{timeepsilonf1} and figure \ref{timeepsilonf2}. The interaction energy is changed from $\epsilon=0.1$ to $\epsilon=8$. The external force is changed from $f=1$ in figure  \ref{timeepsilonf1} to $f=2$ in figure \ref{timeepsilonf2}. As it appears from both figures, increasing the pore diameter will decrease the translocation time. Moreover, while increasing the interaction energy, generally will increase the translocation time, this increase is very small in the low interaction energies ($\epsilon=0.1, 1$). As expected, increasing the external force will increase the translocation velocity. The mean waiting time of each monomer for different pore radii of $1.5\sigma$, $2.5\sigma$ and  $3.0\sigma$ is plotted against the monomer number, s, in figure \ref{waiting1.5}, \ref{waiting2.5} and figure \ref{waiting3.0}, respectively. The maximum of the translocation time is related to the middle monomers due to the entropic barrier of the cis and trans monomers. Thus the mean waiting plots are nearly bell-shape. The behavior of the final monomers in the interaction energy of $\epsilon_{0}=8$ and nanopore of radius $r=1.5\sigma$ is interesting. As it appears in figure \ref{waiting1.5}, the final monomers waiting times for $\epsilon_{0}=8$ and for both external forces of $f=1$ and $f=2$ are ascending. Because of the large interaction energy, the final monomers do not want to come out.

\begin{figure}
  \includegraphics[width=\linewidth]{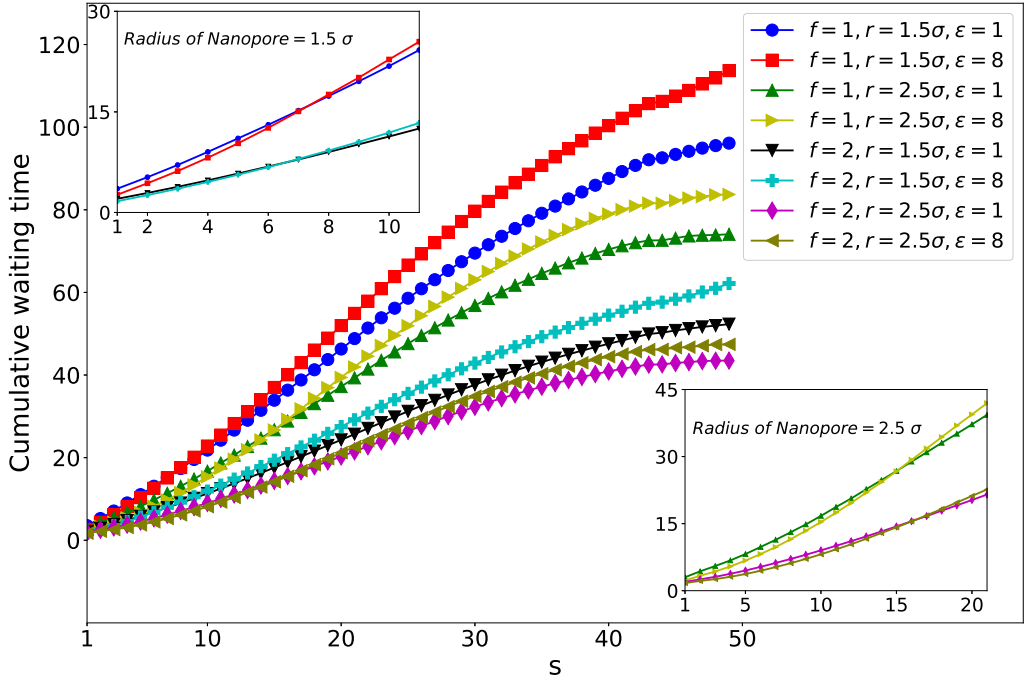}
  \caption{Cumulative waiting times versus monomers number s}
  \label{Cumulative_waiting}
\end{figure}

Cumulative waiting time versus monomer number, s, is shown in figure \ref{Cumulative_waiting}. It compares, different interaction energies of $\epsilon_{0}=1, 8$, different external forces of $f=1, 2$, and different pore radii of $r=1.5\sigma, 2.5\sigma$. Insets are the zoom of the plots at first monomers. As the top inset shows for $r=1.5\sigma$ the polymer with $\epsilon_{0}=8$ is faster than the interaction energy of $\epsilon_{0}=1$ at 6 first monomers for both forces of $f=1, 2$. In the wider pore where $r=2.5\sigma$ the intersection of plots becomes on $s=13$ (see the low inset of the figure \ref{Cumulative_waiting}). It means that the high interaction pulls the polymer through the pore and makes it faster at first, but slows its translocation through the pore in the middle stages. This effect becomes more important as the pore radius becomes larger. We expect that by increasing the radius of the pore till the point where it is still smaller than the gyration radius of the polymer, and also the interaction of the nanopore with the polymer is large enough, this monomer number will rise.

To justify such behaviors in the polymer translocation, we need to look at other parameters such as the center of mass (COM) of the polymer during the passage, the overall shape of the polymer (shape factor) and the spatial distribution of monomers through the translocation process, etc.

\begin{figure}
  \includegraphics[width=\linewidth]{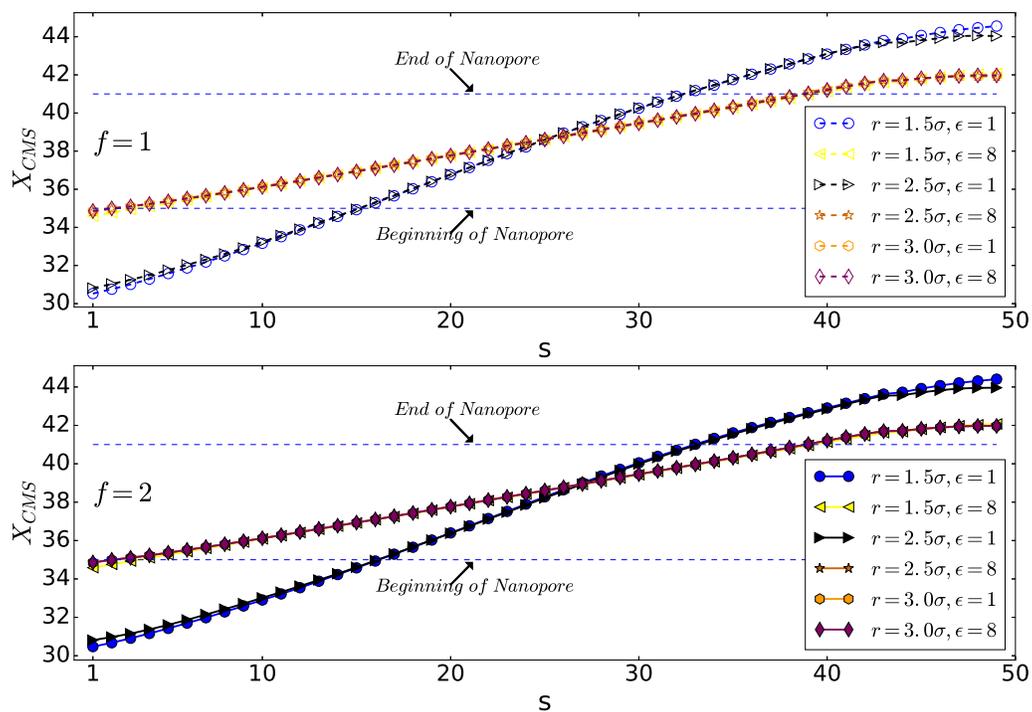}
  \caption{x component of the location of the center of mass (COM) of the polymer versus monomer number, s.}
  \label{xcms}
\end{figure} 


The figures \ref{xcms} show the X components of the COM of the polymer versus monomer's number, s, respectively. The pore center coordinate is (40, 38, 40). It is important to mention here that the polymer is initially in equilibrium. To discuss the translocation in more detail, we focus on $X_{COM}$ which is the pore direction in figure \ref{xcms}. As it shows, in the first stage of the translocation the polymers with high interaction energy of $\epsilon_{0}=8$ have greater $X_{COM}$ from the polymers with low interaction energy of $\epsilon_{0}=1$ which means they reach to equilibrium nearest to the pore as the interaction supports. They are also nearest to the pore in the last stage of the translocation with the same reason. To see the polymer's behavior in more detail, we study the polymer shape using the parameters $\alpha$ and shape factor. $\alpha$ compares the distribution of the monomers in pore axis (x) and from the translocation axis (in yz plane), $\alpha = \Delta x/(2r)$. $\Delta x$ is the maximum of the polymer distance from the pore in the trans side in the x-direction and r is the maximum distance of the polymer from the pore axis (x) in the trans side, $r=\sqrt{y^{2}_{max} + z^{2}_{max}}$ \cite{18Katkar}.

\begin{figure}
  \includegraphics[width=\linewidth]{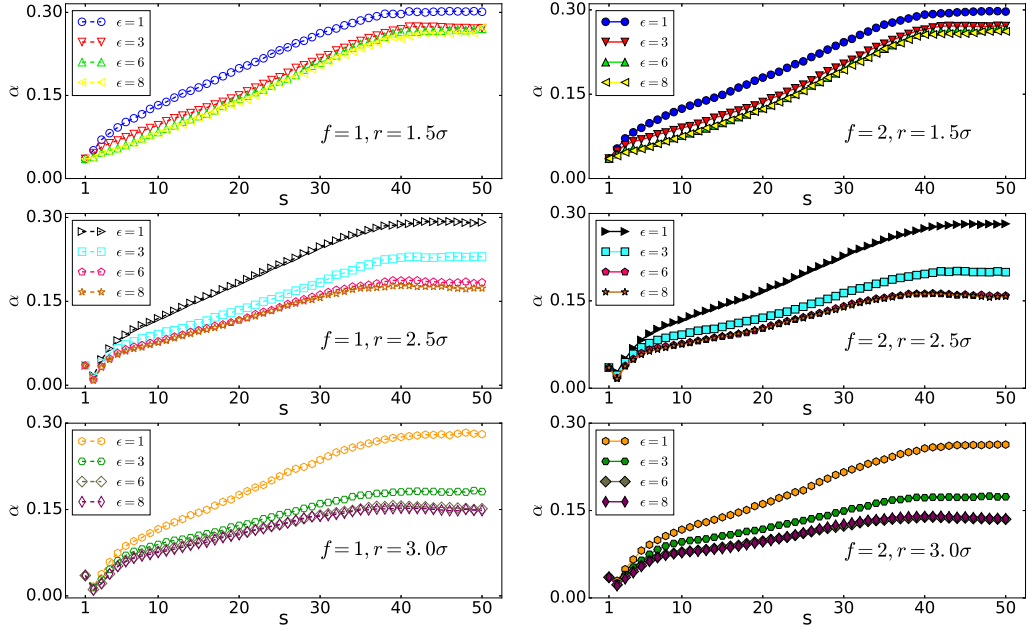}
  \caption{$\alpha$ versus monomer number s.}
  \label{alpha_ep}
\end{figure}

\begin{figure}
  \includegraphics[width=\linewidth]{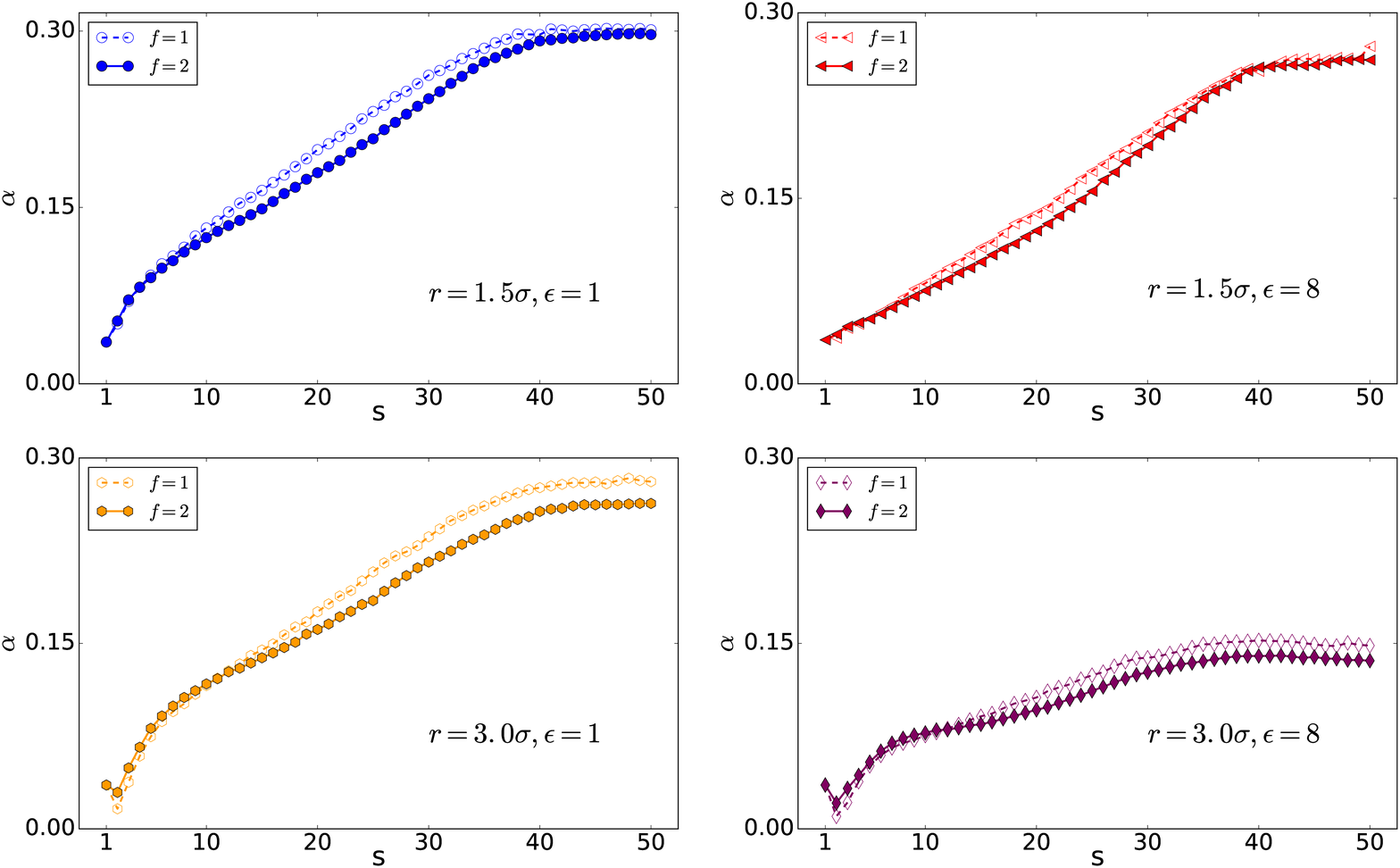}
  \caption{$\alpha$ versus monomer number s.}
  \label{alpha_fo}
\end{figure}

\begin{figure}
  \includegraphics[width=\linewidth]{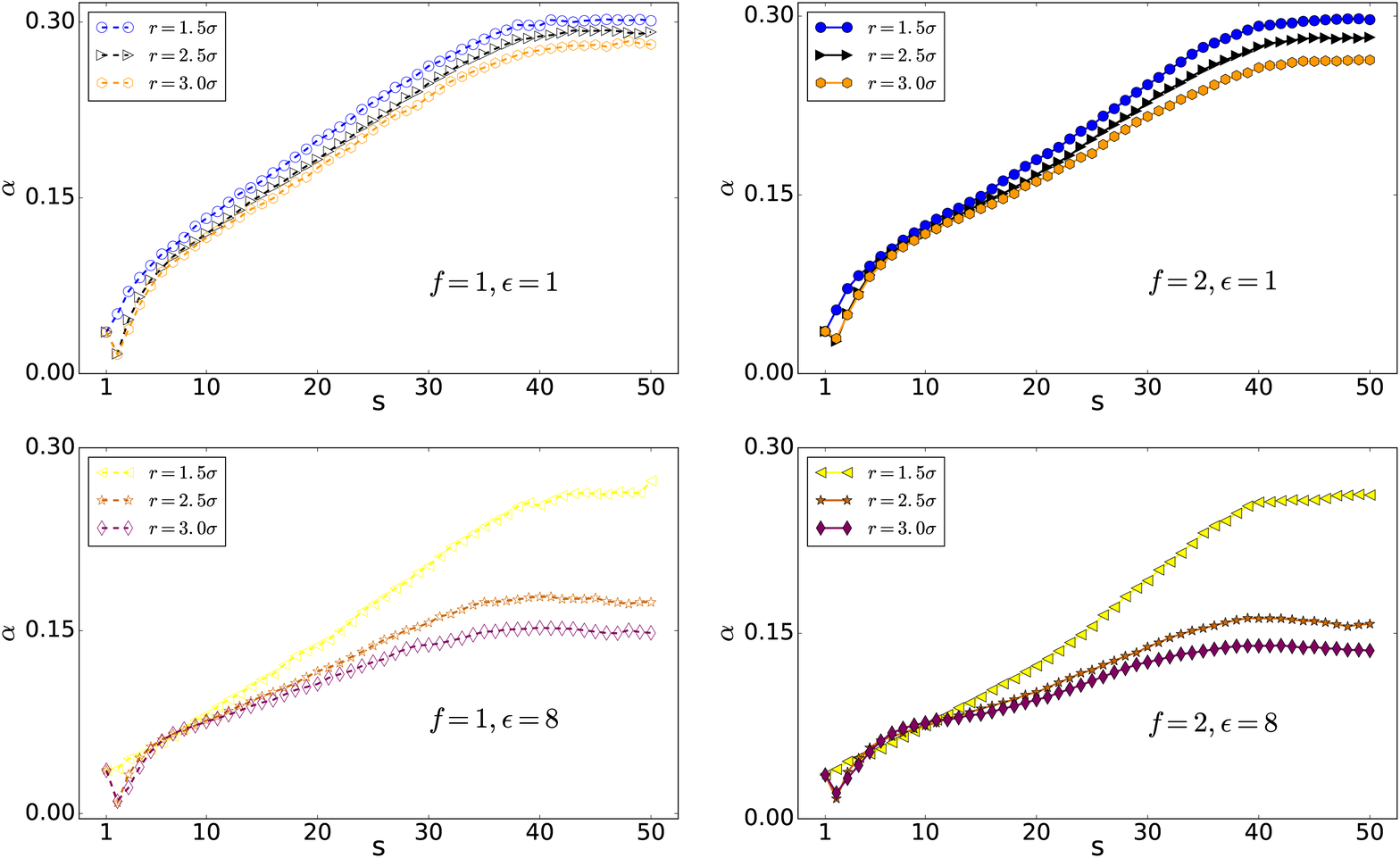}
  \caption{$\alpha$ versus monomer number s.}
  \label{alpha_ra}
\end{figure}

As the figures \ref{alpha_ep}, \ref{alpha_fo} and \ref{alpha_ra} show the distribution of monomers in wider pore of $r=2.5 \sigma$ is thinner than the pore with radius $r=1.5 \sigma$. Moreover, it shows that in accordance to the previous discussion, the narrowest of distribution of the monomers is in the case of high interaction energy of $\epsilon_{0}=8$ and in $r=2.5 \sigma$. 

\begin{figure}
  \includegraphics[width=\linewidth]{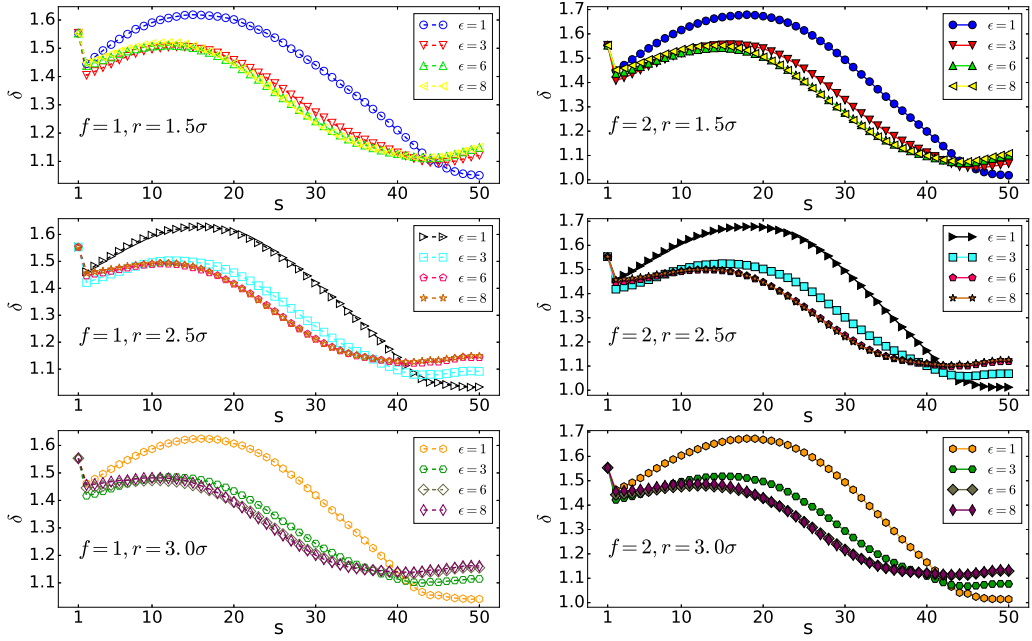}
  \caption{shape factor $\delta$ versus monomer number s}
  \label{delta_ep}
\end{figure}

\begin{figure}
  \includegraphics[width=\linewidth]{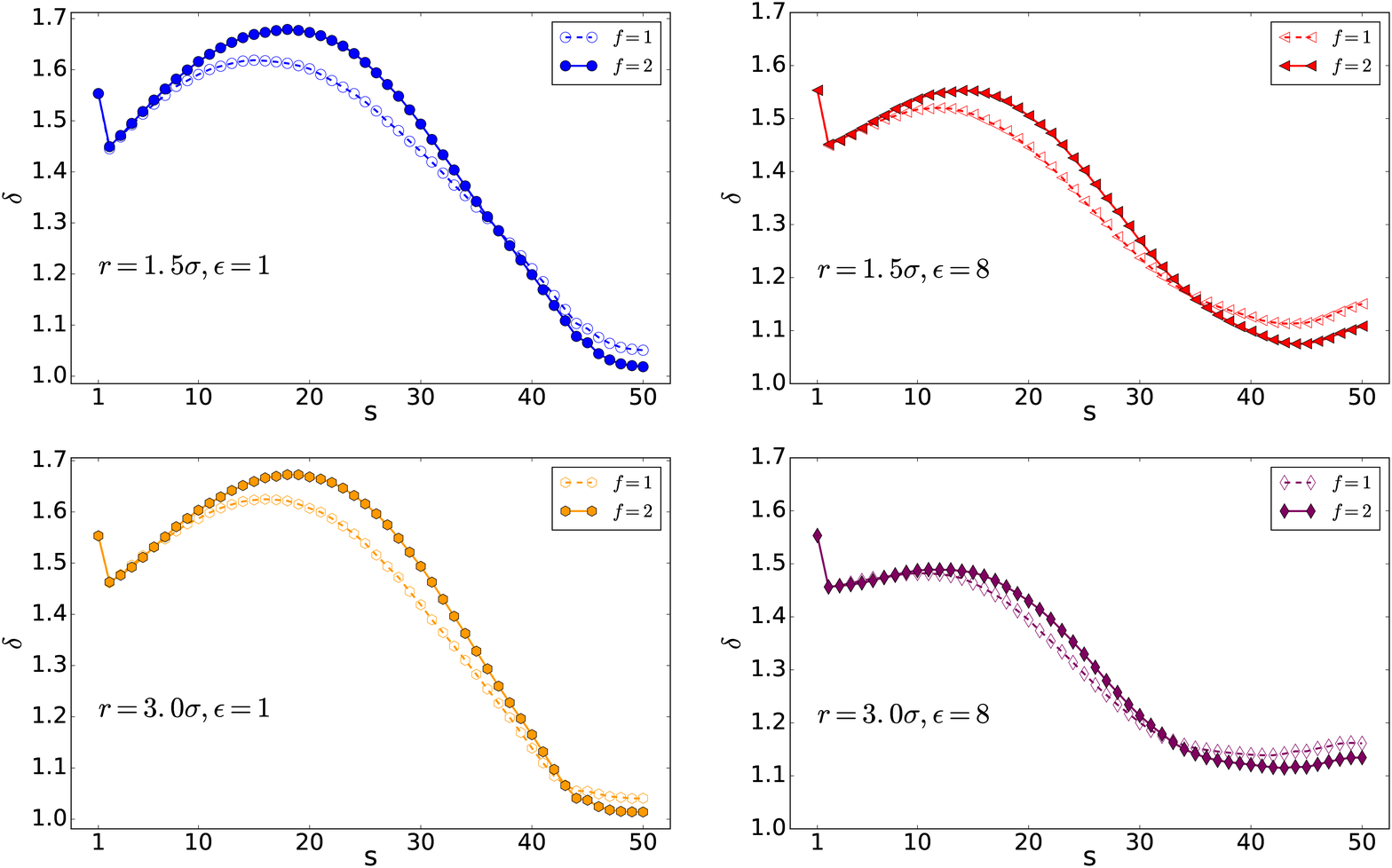}
  \caption{shape factor $\delta$ versus monomer number s}
  \label{delta_fo}
\end{figure}

\begin{figure}
  \includegraphics[width=\linewidth]{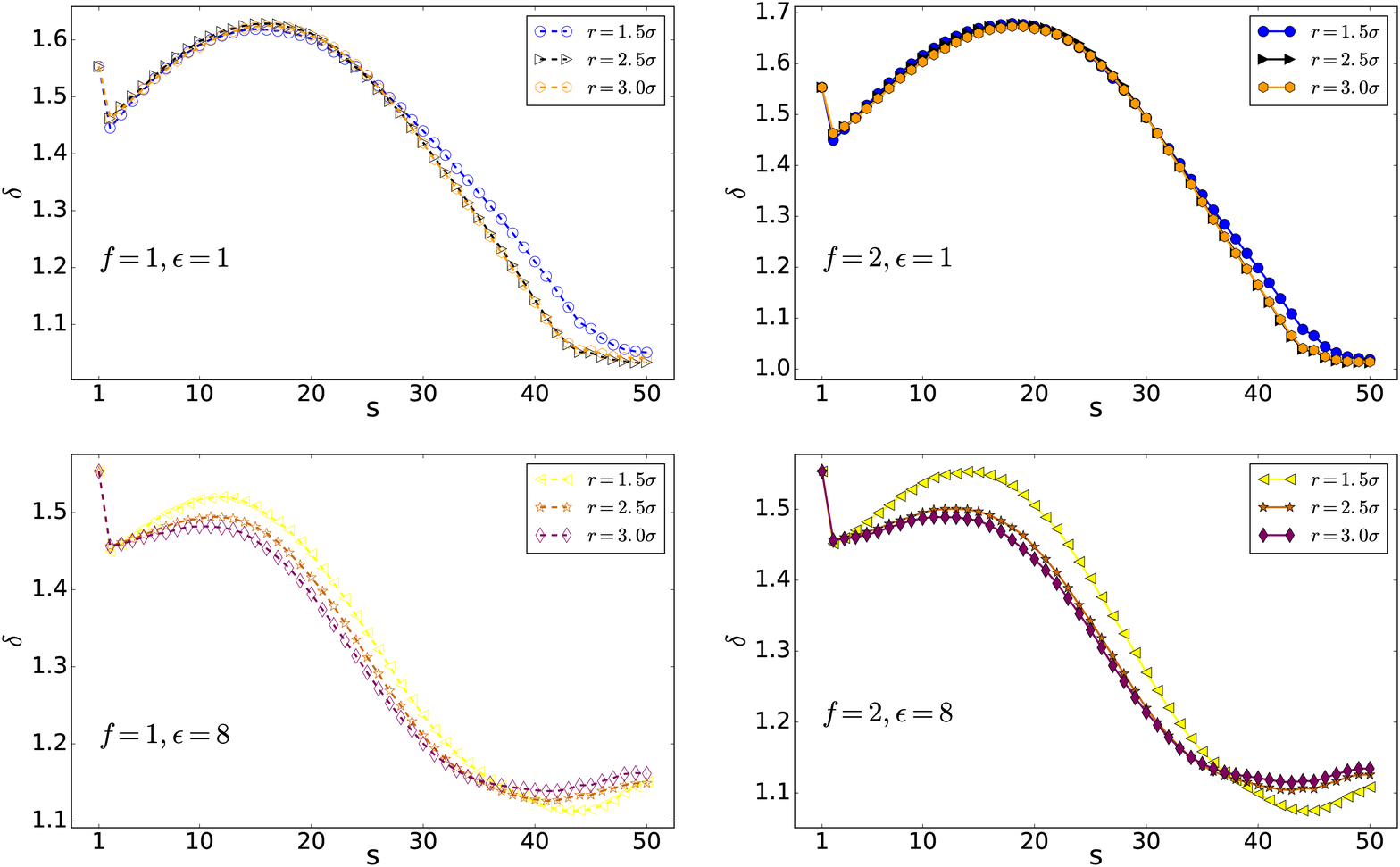}
  \caption{shape factor $\delta$ versus monomer number s}
  \label{delta_ra}
\end{figure}

The shape factor $\delta$ versus monomer number have been shown in figures \ref{delta_ep}, \ref{delta_fo} and \ref{delta_ra}. This parameter compares the gyration radius and the hydrodynamic radius \cite{11muthukumar}. The upper limit of the shape factor $\delta$ is for a rod and equals $\delta_{max}=4.0$ and the lower limit of it is for a compact sphere and equals $\delta=0.77$ \cite{11muthukumar}.

They show that increasing the interaction energy will decrease the shape factor variation. Moreover, increasing the external force $\delta$ will increase, and the polymer becomes more rod shape. In addition, at the final stage of the translocation, the shape factor $\delta$ will increase by increasing the interaction energy. It means that the polymer with lower interaction energies is more compact with respect to those with higher $\epsilon_{0}$.

\section{Conclusions}
\label{conc}

We use a 3D molecular dynamics to simulate the polymer translocation through a narrow pore driven by an external force. Simulation results show that increasing the polymer-pore interaction energy slows down the translocation. Moreover, increasing the pore diameter makes the translocation faster which is in complete accordance with previous results \cite{16Menais,18Menais}. 

The detailed analysis of the polymer shape shows that the polymer wants to be more near the pore in high energies at both first and last part of the translocation process with respect to the polymers with lower interaction energies. This cause the translocation of the high interaction polymers becomes faster at first and slower at last. Moreover, our detailed shape analysis reveals that the polymers with lower energy and in wider pores are more rod shape through the translocation. Also, while the polymer shape is not sensible to the external force (at least in the forces of $f=1$ and $f=2$), its shape is very sensitive to the interaction energy between the polymer and nanopore. 

Waiting time analysis shows that the middle monomers take more time than others. In high interaction energy of $\epsilon_{0}=8$ and the small pore radius of $r=1.5\sigma$, the last monomer's waiting times versus monomer number are ascending. Due to the high interaction and accumulation of the monomers at the trans side, the polymer does not want to come out of the pore. 

In summary, changing the pore diameter and polymer-pore interaction will cause the translocation time, polymer shape through the translocation, accumulation of the monomer at first and last stage of the translocation and waiting time of each monomer to variate widely.

\section{Acknowledgments}
The Molecular Dynamics simulations were performed with the ESPResSo package \cite{espresso1,espresso2,espresso3}. Simulation plotting's been done by using Matplotlib \cite{matplotlib_cite}.

\bibliographystyle{elsarticle-num}
\section*{\refname}
\bibliography{MyReferences970806}

\end{document}